\documentclass[fleqn,usenatbib, letter]{mnras}
\usepackage{amsmath}
\usepackage{mathptmx}
\usepackage{txfonts}
\usepackage{graphicx}
\usepackage[T1]{fontenc}
\usepackage{ae,aecompl}

\usepackage{graphicx}
\usepackage[pdf]{pstricks}
\usepackage{epstopdf}
\usepackage{amssymb}
\usepackage{latexsym}
\usepackage{tikz-cd}
\usepackage{bm}
\usepackage{natbib,twoopt}
\usepackage{subfig}

\newcommand{\dd}{{\rm d}}
\newcommand{\bam}{\texttt{BAM} }

\title[BAM:  Bias Assignment Method for mock catalogs]{BAM: Bias Assignment Method to generate mock catalogs}

\author[Balaguera-Antol\'{\i}nez et al]{\parbox{\textwidth}
  {A. Balaguera-Antol\'{\i}nez\thanks{balaguera@iac.es}$^{1,2}$, Francisco-Shu Kitaura\thanks{fkitaura@iac.es}$^{1,2}$, Marcos Pellejero-Ib{\'a}{\~n}ez$^{1,2}$, Cheng Zhao$^{3}$ and Tom Abel$^{4}$\\
  }
  \\
  $^{1}$Instituto de Astrof\'{\i}sica de Canarias, s/n, E-38205, La Laguna, Tenerife, Spain\\
  $^{2}$Departamento de Astrof\'{\i}sica, Universidad de La Laguna, E-38206, La Laguna, Tenerife, Spain\\
  $^{3}$National Astronomy Observatories, Chinese Academy of Science, Beijing, 100012, P.R. China\\
  $^{4}$Kavli Institute for Particle Astrophysics and Cosmology, Stanford University,
SLAC National Accelerator Laboratory, \\  Menlo Park, California 94025, USA}
\begin{document}
\label{firstpage}
\pagerange{\pageref{firstpage}--\pageref{lastpage}} 
\maketitle

\begin{abstract}
 We present BAM: a novel Bias Assignment Method envisaged to generate mock catalogs. Combining the statistics of dark matter tracers from a high resolution cosmological $N$-body simulation and the dark matter density field calculated from down-sampled initial conditions using efficient structure formation solvers, we extract the halo-bias relation on a mesh of a $3\,h^{-1}$ Mpc cell side resolution as a function of properties of the dark matter density field (e.g. local density, cosmic web type), automatically including  stochastic, deterministic, local and non-local components. We use this information to sample the halo density field, accounting for ignored dependencies through an iterative process. By construction, our approach reaches $\sim 1\%$ accuracy in the majority of the $k$-range up to the Nyquist frequency without systematic deviations for power spectra (about $k \sim 1\, h$ Mpc$^{-1}$) using either particle mesh or Lagrangian perturbation theory based solvers. When using phase-space mapping to compensate the low resolution of the approximate gravity solvers, our method reproduces the bispectra of the reference within $10\%$ precision studying configurations tracing the quasi-nonlinear regime. BAM has the potential to become a standard technique to produce mock halo and galaxy catalogs for future galaxy surveys and cosmological studies being highly accurate, efficient and parameter free. 
\end{abstract}



\section{Introduction}\label{sect1}
The analysis of cosmological large-scale structure experiments such as eBOSS \cite[][]{2016AJ....151...44D}, DES \cite[][]{DES}, Euclid \cite[][]{Euclid} and DESI \cite[][]{DESI} demands exact models of galaxy clustering and precise estimates of covariance matrices \cite[e.g.][]{2013PhRvD..88f3537D,2013MNRAS.432.1928T}. In the lack of analytically accurate models capturing the highly complex nonlinear gravitational evolution underlying a galaxy distribution, the physical processes such as galaxy bias, baryon effects \citep[e.g.][]{1998ApJ...496..605E,2008ApJ...672...19R}, redshift space distortions \citep[e.g.][]{1987MNRAS.227....1K} and systematic effects (e.g. survey geometry), the construction of large sets of mock catalogs based on $N$-body simulations (N-BS) has become  the standard approach to asses robust error estimates on cosmological observables. This imposes some practical restrictions, given the considerably high time and/or memory requirements that a state-of-the-art N-BS requires to generate hundreds to thousands of realizations and light-cones. 
In order to speed-up the construction mock catalogs, a number of alternatives have been designed. Pioneer works such as \texttt{PINOCCHIO} \citep{2002MNRAS.331..587M} and the \texttt{PEAK-PATCH} method \citep{1996ApJS..103....1B} aimed at generating mock galaxy (or halo) catalogues in a predictive way by using approximate gravity solvers based on analytical approaches, such as Lagrangian perturbation theory and prescriptions to compute the formation of halos. The problem of these methods is that such approximations do not properly describe structure formation on small scales, deviating beyond 5\% accuracy from the true power spectrum already on scales relevant to baryon acoustic oscillations, redshift space distortions and non-linear evolution. There are also some recent advances in fast gravity solvers such as \texttt{ICE-COLA} \citep{2013JCAP...06..036T, 2016MNRAS.459.2118K,2016MNRAS.459.2327I} and \texttt{FastPM} \citep{2016MNRAS.463.2273F}, which do not suffer so severely from these inaccuracies.  All these methods however, require resolving the halos and thus demand approximately the same level of memory as a full $N$-body calculation, being just moderately faster in the computational process \citep[e.g][]{2018arXiv180609497B}. These strategies also limit the methods to generate mock galaxy catalogues, as they work best to resolve distinct halos, which can be then augmented with, e.g. an HOD approach \cite[e,g.][]{2003ApJ...593....1B, 2005ApJ...630....1Z}. However, techniques such as Halo Abundance Matching \cite[e.g.][]{2010ApJ...717..379B, 2011ApJ...742...16T} require resolving also the substructures, which can only be obtained with very accurate gravity solvers, and thus remain unreachable for those approaches. Special mention should be made to a different pioneering work, \texttt{PThalos} \citep{2002MNRAS.329..629S,2013MNRAS.428.1036M,2015MNRAS.447..437M},  further explored in later works by \texttt{PATCHY} \citep[][]{2014MNRAS.439L..21K}, \texttt{QPM} \citep{2014MNRAS.437.2594W}, and \texttt{EZMOCKS} \cite[][]{2015MNRAS.446.2621C} and \texttt{HALOGEN} \citep{2015MNRAS.450.1856A}. The idea in these approaches is to rely only on the smooth large-scale dark matter field obtained from approximate gravity solvers, and populate it with halos (or galaxies) following some bias prescriptions. The problem  with this approach is that the bias prescription is not trivial. Many studies have demonstrated that this quantity depends on several properties of dark matter halos, such as their assembling history \cite[e.g.][]{2007MNRAS.377L...5G} their local density and mass \cite[e.g.][]{pillepich_nm,2010ApJ...724..878T,2011A&A...525A..98V}, scale-dependency and non linear evolution \cite[e.g.][]{1993ApJ...413..447F, 1999ApJ...520..437K, 2007PhRvD..75f3512S}, stochasticity \cite[e.g][]{1996MNRAS.282..347M, 1999ApJ...520...24D,2000ApJ...540...62S, 2001MNRAS.320..289S,2014MNRAS.439L..21K}, non-locality \cite[e.g.][]{1999ApJ...525..543M, 2009JCAP...08..020M, 2012PhRvD..85h3509C, 2012MNRAS.420.3469P, 2013PhRvD..87h3002S}, the cosmic web type \citep[][]{2017arXiv170402451Y, 2018MNRAS.473.3941F}, among others \citep[see e.g.][ for recent results]{2018MNRAS.476.3631P}. The halo bias, from a practical implementation, also depends on the chosen resolution of the dark matter mesh, approximations of the gravity solver, and mass assignment scheme (MAS hereafter). In fact this remains hitherto as the main problem for all these bias mapping approaches, requiring complex calibration procedures  \citep[e.g.][]{2015MNRAS.450.1836K,2017MNRAS.472.4144V}.
Despite of such difficulties, these bias mapping methods have shown to reach high level of accuracy circumventing the limitations of the approximate gravity solvers \cite[see][]{2015MNRAS.452..686C} being able to make large amount of precise mock galaxy catalogues \citep[e.g.][]{2016MNRAS.456.4156K} with very low memory and computational requirements \citep[e.g][]{2018MNRAS.tmp.2818C, 2018arXiv180609497B,2019MNRAS.482.1786L}.
 
Future galaxy surveys will trace the cosmic web further towards the low density regime, being able to map in more detail the filamentary network with bright, blue, red, and emission-line galaxies \citep[e.g.][]{2016A&A...592A.121C, 2018MNRAS.474..177M,fmost,2014arXiv1403.5237B}. In the light of this situation, and with a calibration process which is becoming too complex and subject to propagation of systematic errors due to an approximate bias modelling, we propose here to radically simplify the procedure trying to capture the full complexity, by directly mapping the bias relation from detailed N-BS. To this aim, in this letter we present a new method to generate mock catalogs: the Bias Assignment Method (\texttt{BAM}), which applies the idea of halo bias mapping in a free-parameter fashion. To describe the nonlinear dark matter field accurately with low number of particles we resort on the phase-space mapping technique (PSM hereafter) \citep[][]{2012MNRAS.427...61A,Hahn:2013aa}. In order to develop our method we adopted the \texttt{MINERVA} simulations \citep{2016MNRAS.457.1577G} with output at redshift $z=1$. The simulation has a comoving volume of $1500^{3}$ $({\rm Mpc}/h)^{3}$ with a mass resolution of $M_{\rm min}=2.6\times 10^{12}$ $M_{\odot}$ and $1000^{3}$ dark matter (DM hereafter) particles. Dark matter halos (DMH hereafter) are identified with a \emph{friends-of-friends} algorithm (FoF hereafter).
The outline of this letter is as follows. We describe the \bam assumptions and method and study its performance. Then we end-up with conclusions and discussion.

\begin{figure}
  \vspace{-0.5cm}
   \hspace{-0.5cm}
  \includegraphics[width=9.5cm, height=7.8cm]{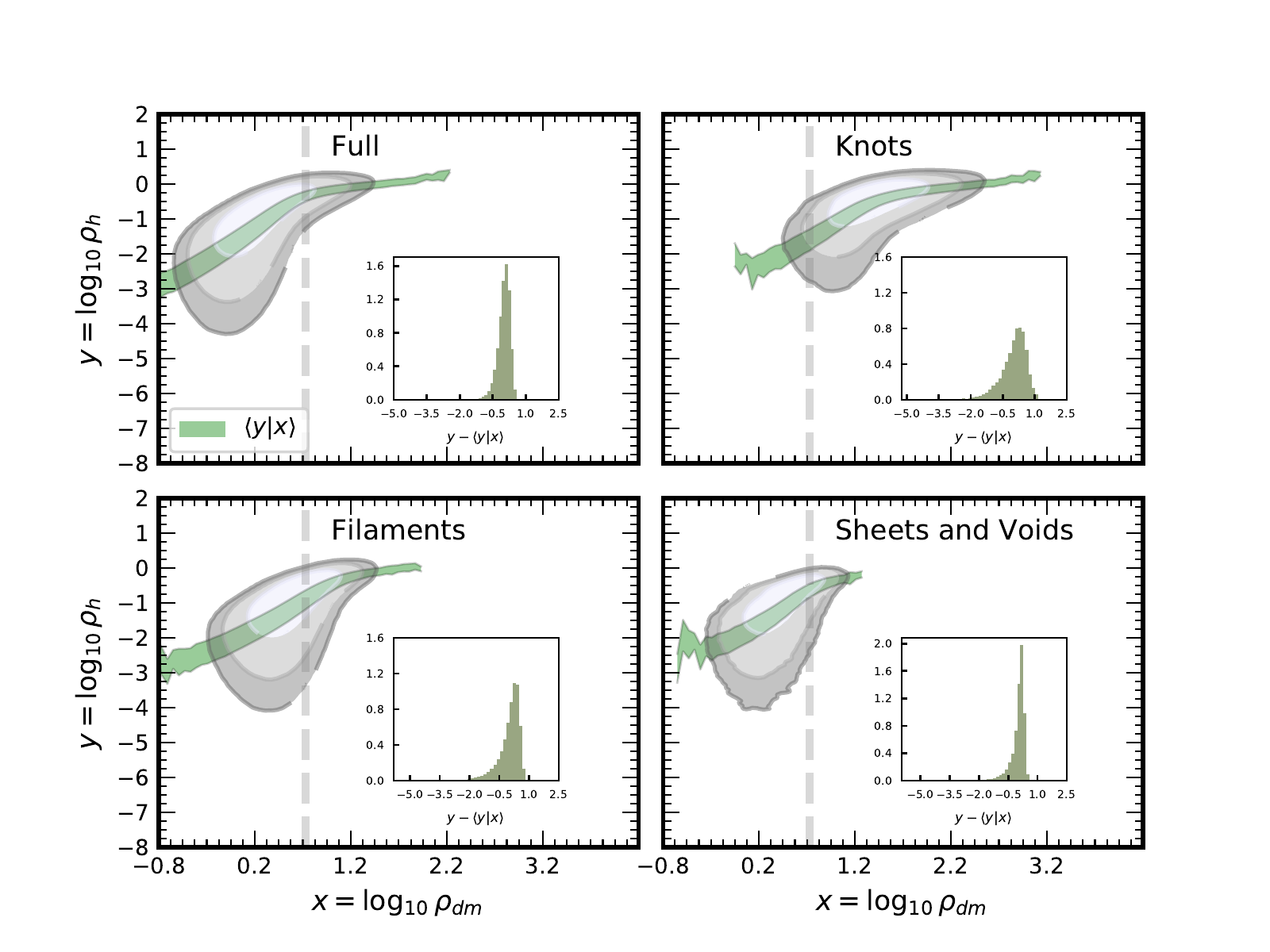}
  \vspace{-0.5cm}
\caption{Halo bias for different cosmic web classifications, in the plane $x=\log (\rho_{\rm dm})$  and $y=\log{\rho_{\rm h}}$, where $\rho=1+\delta$, is the density interpolated on the grid. The contours in each panel denote the region containing $65$, $95$ and $99\%$ of the total number of classified cells. The filled curve denotes the mean relation $\langle y| x \rangle$ and its \emph{rms}, while the insets display one element of the full $P(y|x)\dd x$, where the bin $(x, x+\dd x)$ is represented by the vertical line.} 
  \label{bias_bam}
\end{figure}

\section{The \bam method}\label{sec:bam}
\bam exploits the idea of mapping the halo distribution onto a target dark matter density field (TDMF hereafter) with a biasing scheme extracted from the DMH distribution from a reference N-BS and a TDMF obtained by evolving approximate gravity solvers using the same initial conditions of the reference simulation, downgraded to a lower resolution. 
Such mapping can then be applied to any configuration of initial conditions retaining the same cosmological parameters and numerical setting.
This builds on the rank ordering method proposed by \citet[][]{doi:10.1093/mnras/254.2.315}, extending it to a multivariate bias relation dependent on local and non-local quantities \citep[e.g.][]{2018arXiv180209177H}, and relating a continuous field to a discrete realization of tracers. Moreover, our approach includes an iterative sampling procedure accounting for cross-correlations and other dependencies ignored in the analysis. 

\begin{figure*}
  \vspace{-1cm}
  \includegraphics[width=17cm]{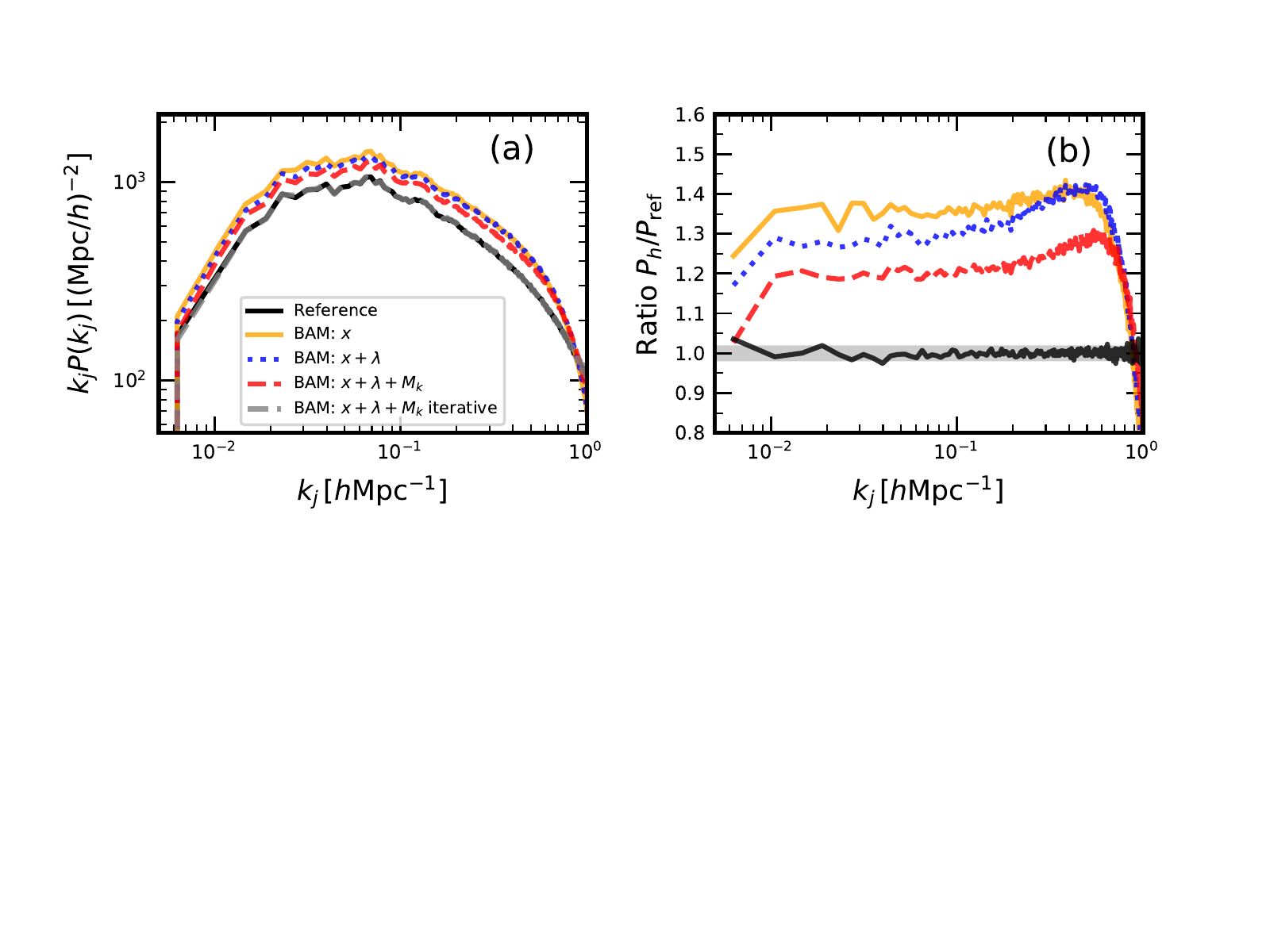}
  \vspace{-6cm}
  \caption{Panel (a): the solid (black) line represents the power spectrum of the reference halo catalog. Solid (orange) line represents the power spectrum of a \bam mock using the information from the local density. Dotted (blue) line includes the information from the cosmic-web, while the dashed (red) includes the environmental dependencies. The thick dashed (grey) line represents the results from the iterative process described in the text. Panel (b) shows the ratio if the spectra shown in panel (a) to the reference power spectrum. The solid (black) line represents the results from the iterative approach of \bam. The shaded area indicates $2\%$ deviations around unity, with the majority of the modes lying within $1\%$ up to $k\sim0.6\,h\,{\rm Mpc}^{-1}$.}\label{power_bam}
\end{figure*}

We characterize the halo bias in the spirit of \cite{1999ApJ...520...24D}, measuring the probability distribution of halo number densities conditional to a set of properties of the underlying DM density field such as \emph{the local density}, the \emph{the cosmic web-type} and \emph{environmental density}. Given that the properties of DMH are expected to be tightly correlated to the DM density field \citep[e.g.][]{1986ApJ...304...15B}, our method
indirectly takes into account the dependencies of the halo bias on halo properties of the target population.
To motivate the basis of \texttt{BAM}, in the top-left panel of Fig.~\ref{bias_bam} we show with shaded contours the joint probability distribution $P(x,y;\Delta V)$ from the reference simulation, measured in cells of volume $\Delta V=9({\rm Mpc}\,h^{-1})^{3}$, where $x=\log_{10}(1+\delta_{\rm DM})$ and $y=\log_{10}(1+\delta_{\rm H})$, with $\delta$ denoting overdensities. For illustrative purposes, we used the CIC mass assignment interpolating the DM and DMH onto the mesh. The shaded curve denotes the first two moments of the conditional probability distribution $P(y|x)$ (i.e., mean and \emph{rms}). By comparing the shaded curve and the contours, it is evident that there is more information on halo bias beyond the first and the second moments of the conditional probability distribution. 
The vertical dashed line in the same panel represents a bin of DM density, in which the distribution of the DMH densities behaves as shown in the inset plots.

In order to characterize the halo bias as a function of the cosmic web type (denoted by $\lambda$), we determine in the mesh the eigenvalues of the tidal field tensor of the DM distribution, and for a threshold value $\lambda_{\rm th}=0$, we classify DMH as knots, filaments, sheets, or voids \cite[see e.g.][]{2007MNRAS.375..489H}. Top-right and bottom panels of Fig.\ref{bias_bam} represents the joint probability distribution $P(y,x;\Delta V)_{\lambda}$ and its first two moments, for different cosmic-web types. As for the first panel, we show in the inset an example of the conditional distribution $P(y|x;\Delta V)_{\lambda}$, in order to evidence the difference in the halo bias in each case. Finally, we account for the halo environment \citep[see e.g.][and references therein]{2018MNRAS.473.2486S}, defined here as large-scale collapsing regions identified as the percolation of cells classified as knots, by means of a FoF algorithm \citep{2015MNRAS.451.4266Z}. We compute the mass $M_{K}$ of such regions from the dark matter particles therein contained, labeling each associated cell with that value of $M_{K}$. Our method can be extended to other properties of the DM density field.

\begin{figure*}
  \includegraphics[width=16cm, height=12.5cm]{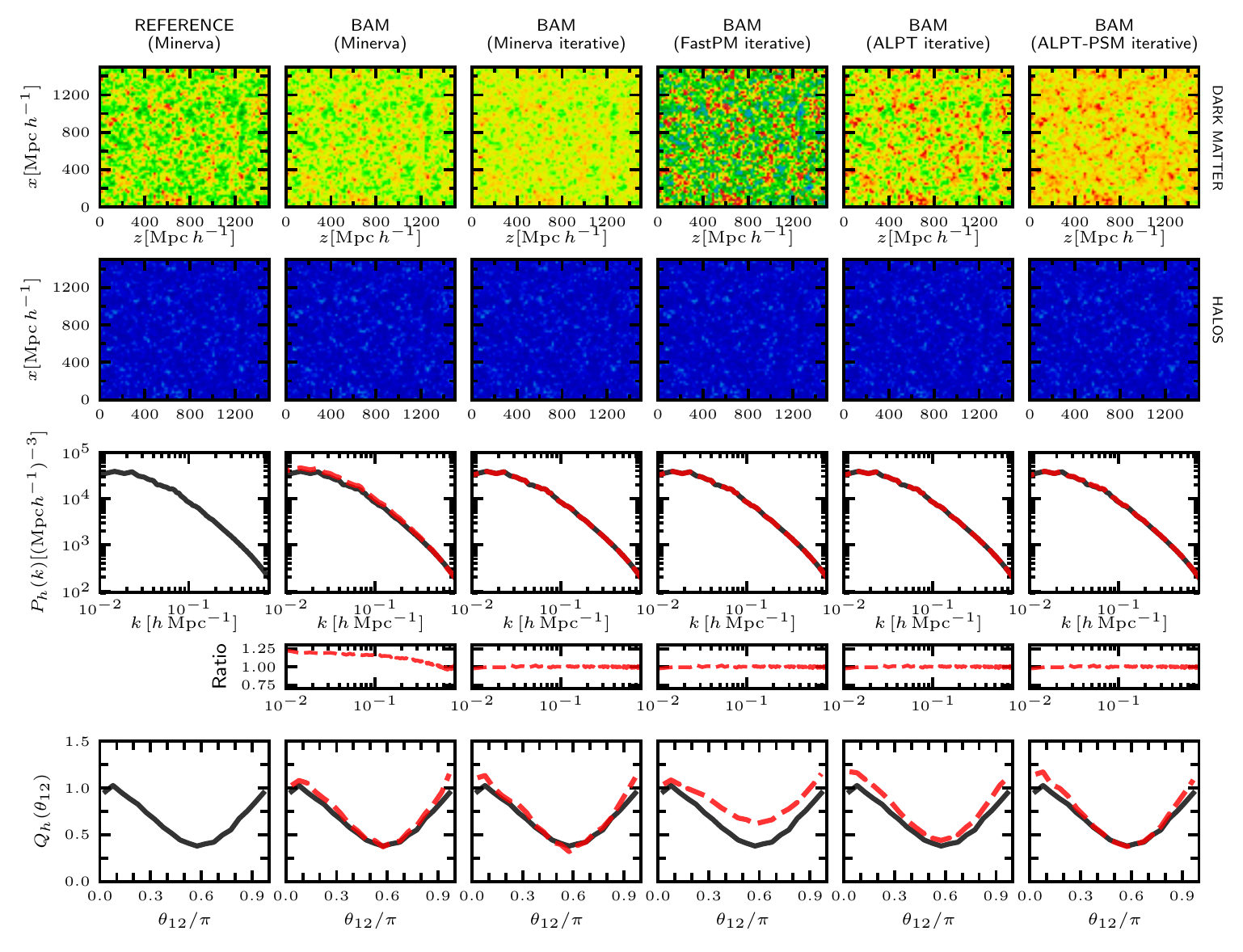}
 \hspace{-0.5cm}
  \includegraphics[width=0.7cm, height=12.25cm]{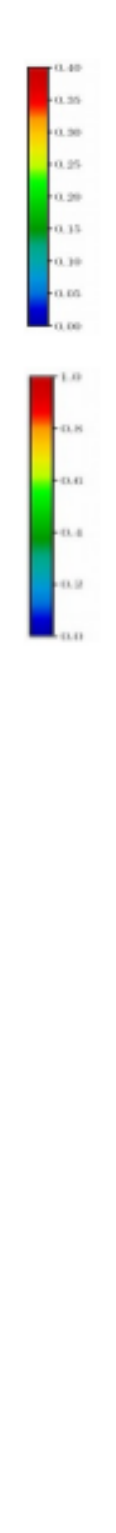}
\vspace{-0.5cm}
\caption{The first and second rows represents a slice of width $\sim 20$ Mpc of the dark matter halo density fields, respectively. We show the reference simulation (Minerva, first column), the density fields after one (second column) and several (third column) iterations with \bam, using the Minerva DM as TDMF. Fourth and fifth columns show the results of using \bam with two approximated gravity solvers, namely, \texttt{ALPT} and \texttt{FastPM}. The last column shows \texttt{ALPT} combined with PSM. The color scale is the same for all dark matter (or halo) density fields, and represents $\log_{10}(2+\delta)$. The third row shows the corresponding halo power spectrum compared to that obtained in each case (dotted line), while the the fourth row shows their ratios to the power spectrum of the reference. The last row shows the reduced bispectrum $Q(k_{1},k_{2},\theta_{12})$, measured for a configuration $k_{1}=0.1\,h$ Mpc$^{-1}$ and $k_{2}=2k_{1}$.}
  \label{dens_bam}
\end{figure*}

The main inputs required to generate a mock halo number density field are $i)$ the reference halo number counts in cells, obtained from the halo catalog constructed from the DM particle distribution, and $ii)$ the TDMF, obtained from an approximated gravity solver using \emph{the same initial conditions of the reference simulation}. With these inputs, the steps followed by \bam are summarized as
\begin{itemize}
    \item Classification of the cosmic web based on the TDMF.
    \item Identification of the large-scale collapsing regions ($M_{K}$).
    \item Measurement of the halo bias, i.e., the conditional probability distribution $P(N_{H}|x,M_{K};\Delta V)_{\lambda}\dd x \dd M_{k}$, accounting for the number of DMH from the reference ($N_{H}$) in a cell of volume $\Delta V$ with local DM density in the range $(x,x+\dd x)$, cosmic web type $\lambda$ and embedded in a large-scale collapsing region with mass in the range $(M_{K}, M_{K}+\dd M_{K})$.
    \item Sampling of the TDMF by assigning the number of halos to cells as $N_{H}\frown P(N_{H}|x,M_{K};\Delta V)_{\lambda}$. The procedure is performed such that the distribution of number counts
    replicates that of the reference halo catalog.
\end{itemize}
In order to verify the accuracy of our procedure, we first use the DM distribution of the reference as the TDMF.
In panel (a) of Fig.~\ref{power_bam} we show the power spectrum of the mock catalog obtained after these steps. All power spectra include a Poisson shot-noise correction. To highlight the dependencies described above, we show three cases, namely, $i)$ halo-bias depends solely on the DM density, $ii)$ extending that dependency with the cosmic web type, and $iii)$ including the environment. Panel (b) of the same figure shows the ratio of the power spectrum obtained in each case to the reference halo power spectrum. Assuming that the halo bias only depends on the local DM density, the mock DMH field generated by \bam displays $\sim 40\%$ more clustering than the reference. This fraction is reduced to $\sim 30\%$ when the information of the cosmic web is included, while accounting for the environment lowers it to $\sim 20\%$. This offset is due in part to dependencies of the halo bias not accounted for in our analysis, together with the impact of the mass assignment scheme. \footnote{We have performed some tests on our method: given the DM density field of the reference, interpolated into the mesh with a particular MAS, we can construct a fake reference halo density field using a certain bias prescription (e.g. $N_{H}\frown {\rm Poisson}(\rho_{\rm dm})$). If \bam uses the same MAS for the TDMF  \emph{and} include all relevant properties assumed by the fake halo bias, then our method is able to create a mock halo density field with the power spectrum of the reference with a $<5\%$ precision.} 
In order to account for such unknown dependencies, we introduce an iterative process summarized as follows:

\begin{enumerate}
    \item  Iteration $i=0$: obtain the first DMH field with the steps previously described. Measure its power spectrum $P_{0,j}\equiv P_{i=0}(k_{j})$, where $k_{j}$ denotes the $j-$th spherical shell in Fourier space. Define an isotropic kernel $\mathcal{K}_{ij}$ in Fourier space, initialized to $\mathcal{K}_{0,j}\equiv \mathcal{K}_{0}(k_{j})=1\,\forall j$. 
    \item Iteration $i=1$:  Define a \emph{bias transfer function (BTF)} $T_{1,j}\equiv P_{j,\rm ref}/P_{0,j}$ and assign it to the kernel $T_{1,j} \to \mathcal{K}_{1,j}$. The values of the BTF at each $k_{j}$ assigned to the kernel are selected with a Metropolis-Hasting (MH) algorithm, by computing a transition probability ${\rm min}(1, {\rm exp}(\mathcal{H}^{2}_{0j}-\mathcal{H}^{2}_{1j}))$ with $\mathcal{H}_{ij}=(P_{j,\rm ref}-P_{i,j})/\sigma_{j}$, $\sigma_{j}$ the (Gaussian) variance associated to the reference power spectrum. When the MH criteria does not accept a particular $T_{ij}$, \bam assigns to the kernel its previous value $\mathcal{K}_{i-1,j}$  (unity, for this iteration). 
    
    \item Convolve the TDMF with the kernel $\mathcal{K}$. This generates a new DMF from which a new halo-bias relation is measured. Sample from such bias relation to generate a new DMH density field and estimate the mock power spectrum $P_{1,j}$. 
    
    \item Iteration $i$. Get the BTF $T_{i,j}\equiv P_{j,\rm ref}/P_{i-1,j}$ and update the kernel as $\mathcal{K}_{ij}=T_{1,j}\times  \cdots \times T_{i-1,j}\times T_{i,j}$, where the BTF at each iteration is selected according to the MH algorithm as described in step ii). Repeat step iii) until the spectra $P_{j,\rm ref}$ and $P_{i,j}$ are in agreement to a percent accuracy. 
\end{enumerate}
The black solid line in panel (a) of Fig.~\ref{power_bam} shows the mock halo power spectrum obtained after this process, with its ratio to the reference shown as a black solid line in panel (b). By construction, our method can recover the reference power spectrum with $\sim 1\%$ in the majority of the $k$-range up to the Nyquist frequency without systematic deviations and with few ($\sim 20$) iterations. We have verified that this iterative procedure generates the same precision in the power spectrum with approximately the same number of iterations in the case in which the bias is only explicitly measured as a function of the DM density. We want to stress that, by explicitly using the DM properties described before, we aim at encapsulating all possible unknown dependencies on the halo bias within the \bam kernel. 

The main advantage of \bam can be recognized when applied to an approximated gravity solver, capable to generate dark matter density fields with the correct large-scale structure and with low computational costs. We have explored two examples, the Augmented Lagrangian perturbation theory \citep[\texttt{ALPT},][]{doi:10.1093/mnrasl/slt101} and \texttt{FastPM} \citep{2016MNRAS.463.2273F}, both run using the initial conditions of the reference N-BS. In Fig.~\ref{dens_bam} we summarize the results of these applications. The first (upper) and second row shows the DM and DMH density fields in different stages of \bam. We also show the power spectrum of each case and its ratio to the reference power. The last row in that figure also shows the reduced bispectrum $Q(k_{1},k_{2},\theta_{12})$ for configurations of $k_1=0.1\,h\,{\rm Mpc}^{-1}$ and $k_2=0.2\,h\,{\rm Mpc}^{-1}$, which trace the quasi-nonlinear regime.  
\bam can recover the shape of the latter statistics to a good degree, although with a systematic deviation of $10-30\%$ with respect to the reference. This discrepancy is alleviated using the PSM technique. We have applied it to \texttt{ALPT}, considerably improving the agreement with respect to the bispectrum of the reference catalog, as shown in the fifth column of Fig.~\ref{dens_bam}. We highlight that \emph{the $3$-point statistics is not calibrated in our method}, and still, \bam is able to reproduce the reference bispectrum within a $\sim 10\%$ difference. Whether such deviation is systematic or not will be studied with the large set of Minerva simulations. Further investigation needs to be done to study whether including more information in the halo-bias can reduce these discrepancies. 

\section{Conclusions and discussion}
In this letter we have described \texttt{BAM}, a novel Bias Assignment Method to produce mock halo and galaxy catalogs for future galaxy surveys and cosmological studies. In particular we have demonstrated that \bam reproduces the distribution of halos from an N-BS (the Minerva simulation) within $1\%$ precision in the power spectrum up to Nyquist frequencies of $k\sim 1$ $h$ Mpc$^{-1}$ and within $10\%$ in the bispectrum (using only $500^3$ particles to describe the dark matter field and counts in cell in the $2$- and $3$-point statistics computations). We tested a particle mesh code (\texttt{FastPM}) and a Lagrangian perturbation theory solver (\texttt{ALPT}), performing equally well in the power spectrum. High accuracy in the bispectrum for configurations down to the quasi-nonlinear regime require PSM phase-space mapping, although configurations on very large scales seem to be reproduced equally well without that particular technique. In order to study this more rigorously we plan to use the large set of Minerva simulations. The \bam method is not limited to a particular halo mass to achieve percentage accuracy in the power spectrum, as it does so by construction. Its performance in the bispectrum going to lower masses as well as its application to another realization of the same initial conditions of the reference simulation is currently being investigated (Pellejero-Ib{\'a}{\~n}ez et al, Balaguera-Antol\'{\i}nez et al, in preparation).
It is in any case remarkable how our bias assignment sampling scheme and phase-space mapping enables an efficient Lagrangian perturbation theory solver to reach such high accuracies. Our studies with the \bam method have been restricted to halo populations, to real-space and the computation of the different statistics to number counts per cell, so far. 
The next step, being currently developed, is the assignment of position and velocities to the haloes in each cell and generalize it to mock galaxy catalogs by taking the proper reference catalogs \citep[][]{2016MNRAS.456.4156K}. An improved sub-grid model describing the distribution of objects on small scales still needs to be implemented, able to properly account for small scale effects, such as fiber collision \citep{2017MNRAS.467.1940H}. We note that \bam replaces the parametrized deterministic and stochastic bias prescriptions of the \texttt{PATCHY} method. This new approach represents a big step towards the efficient production of accurate mock catalogs for the analysis of future galaxy surveys.

\section*{Acknowledgements}
We acknowledge the anonymous referee for his/her careful reading of the article and valuable comments that helped to improve the presentation of our work. We also acknowledge Ariel S\'anchez and Claudio Dalla Vecchia for providing us with a realization of the Minerva simulation. ABA acknowledges financial support from the Spanish Ministry of Economy and Competitiveness (MINECO) under the Severo Ochoa program SEV-2015-0548. FSK thanks support from the grants RYC2015-18693, SEV-2015-0548 and AYA2017-89891-P.
MPI acknowledges support from MINECO under the grant AYA2012-39702-C02-01.


\bibliographystyle{mnras}
\bibliography{refs}  
\end{document}